# String Evolution with Friction

**C. J. A. P. Martins*** *and* **E. P. S. Shellard**

*Department of Applied Mathematics and Theoretical Physics*
*University of Cambridge*
*Silver Street, Cambridge CB3 9EW*†

## Abstract

We study the effects of friction on the scaling evolution of string networks in condensed matter and cosmological contexts. We derive a generalized 'one-scale' model with the string correlation length $L$ and velocity $v$ as dynamical variables. In non-relativistic systems, we obtain a well-known $L \propto t^{1/2}$ law, showing that loop production is important. For electroweak cosmic strings, we show transient damped epoch scaling with $L \propto t^{5/4}$ (or, in the matter era, $L \propto t^{3/2}$). A low initial density implies an earlier period with $L \propto t^{1/2}$. For GUT strings, the approach to linear scaling $L \propto t$ is faster than previously estimated.

## 1 Introduction

The scaling evolution of vortex-string networks has been extensively studied analytically in both condensed matter and cosmological settings [1,2,3], but using rather different methods. Condensed matter descriptions tend to focus on a coarse-grained order parameter $\phi$, providing a low-level picture of defect motion by estimating energy dissipation rates. However, high energy physicists take an 'idealized' one-dimensional view of string dynamics by integrating out the radial degrees of freedom (in the Higgs $\phi$ and other fields) to obtain a low-energy effective action—the Nambu action. The resulting equations of motion can then be averaged in order to describe the large-scale evolution of a string network in a so-called 'one-scale' model [1]. This also involves incorporating energy loss mechanisms, including loop production (usually ignored in condensed matter models) and here we add frictional forces due to particle–string scattering. Our generalization of this simple 'one-scale' model allows us to describe the scaling evolution of a string network in dramatically different regimes, ranging from a string-forming phase transition in the early universe through to a rapid quench in a liquid crystal. (Unless otherwise stated, we use units in which $\hbar = c = k_B = 1$.)

* Also at C. A. U. P., Rua do Campo Alegre 823, 4100 Porto, Portugal

† *Email*: cjapm10@damtp.cam.ac.uk and epss@damtp.cam.ac.uk

## 2 The evolution equations

The motion of a cosmic string passing through a background radiation fluid is retarded by particle scattering. For a string resulting from the breaking of a local gauge symmetry the main contribution comes from Aharonov-Bohm scattering, and the corresponding frictional force per unit length is [4]

$$\mathbf{F}_{\mathrm{f}} = -\frac{\mu}{\ell_{\mathrm{f}}}\frac{\mathbf{v}}{\sqrt{1-v^2}}\,, \qquad \ell_{\mathrm{f}} \equiv \frac{\mu}{\beta T^3}\,, \tag{1}$$

where $\ell_{\mathrm{f}}$ is the 'friction lengthscale', $\mathbf{v}$ is the string velocity, $\mu$ is the string energy per unit length, and $T$ is the background temperature. Here, $\beta$ is a numerical factor related to the number of particle species interacting with the string (expected to be of order unity for minimal-GUT strings). The phenomenological force law for a condensed matter system will take a similar form with the same velocity dependence.

In an FRW universe with line element $ds^2 = a^2(\tau)\left(d\tau^2 - \mathbf{dx}^2\right)$, the Nambu string equations of motion with the frictional force (1) can be written [4,5]

$$\ddot{\mathbf{x}} + \left(2\frac{\dot{a}}{a} + \frac{a}{\ell_{\mathrm{f}}}\right)\left(1-\dot{\mathbf{x}}^2\right)\dot{\mathbf{x}} = \frac{1}{\epsilon}\left(\frac{\mathbf{x}'}{\epsilon}\right)'\,, \tag{2}$$

$$\dot{\epsilon} + \left(2\frac{\dot{a}}{a} + \frac{a}{\ell_{\mathrm{f}}}\right)\dot{\mathbf{x}}^2\epsilon = 0\,, \tag{3}$$

where the 'coordinate energy per unit length' is defined by $\epsilon^2 = \mathbf{x}'^2/(1-\dot{\mathbf{x}}^2)$, with dots denoting derivatives with respect to $\tau$, which is identified with the worldsheet time, and primes for derivatives with respect to $\sigma$, a spacelike coordinate labelling points along the string. In deriving (2) and the constraint (3), we require the additional gauge condition that the string velocity be orthogonal to the string direction, that is, $\dot{\mathbf{x}}\cdot\mathbf{x}' = 0$. This proves to be a particularly useful form because dissipation is naturally incorporated in the decay of the coordinate energy density $\epsilon$, while preserving the gauge conditions.

The total string energy is given by $E = \mu a(\tau)\int \epsilon d\sigma$, and we define the average rms string velocity $v$ as $v^2 \equiv \langle\dot{\mathbf{x}}^2\rangle = \int \dot{\mathbf{x}}^2\epsilon d\sigma / \int \epsilon d\sigma$. Then the total string energy density $\rho \propto E/a^3$ will obey the following evolution equation (in terms of physical time $t$, $dt/d\tau = a$):

$$\frac{d\rho}{dt} + \left(2H\left(1+v^2\right) + \frac{v^2}{\ell_{\mathrm{f}}}\right)\rho = 0\,. \tag{4}$$

### The one-scale model

Our aim is to study the evolution of the long-string network on the assumption that it can be characterized by a single lengthscale $L$. We therefore draw a distinction between long strings and small loops according to whether they are, respectively, larger or smaller than $L$. For Brownian long strings, we can define the 'correlation length' $L$ in terms of the network density $\rho_\infty$ as $\rho_\infty \equiv \mu/L^2$.

Following Kibble [1], the probability of a segment of length $l$ moving with a velocity $v$ intercommuting within a time $\delta t$ is approximately $lv\delta t/L^2$. Consistent with our scaling assumption, we then assume that the probability of such an intersection creating a loop of length in the range $l$ to $l + dl$ will be given by a scale-invariant function $w(l/L)$. The rate of energy loss into loops is then given by

$$\left(\frac{d\rho_\infty}{dt}\right)_{\text{to loops}} = \frac{v}{L}\rho_\infty \int w\left(\frac{l}{L}\right) \frac{l}{L}\frac{dl}{L} \equiv \tilde{c}v\frac{\rho_\infty}{L}\,, \tag{5}$$

where the loop 'chopping' efficiency $\tilde{c}$ is assumed to be constant. (Note that $v$ now denotes the average velocity of the long strings only; in previous analyses without friction, this velocity was also assumed to be constant and absorbed into the definition of $\tilde{c}$.) Subtracting the loop energy losses we easily obtain the overall evolution equation for the characteristic lengthscale $L$,

$$2\frac{dL}{dt} = 2HL(1+v^2) + \frac{Lv^2}{\ell_{\mathrm{f}}} + \tilde{c}v\,. \tag{6}$$

### Velocity evolution

We now consider the evolution of the average long-string velocity $v$. A non-relativistic equation can be easily obtained: it is just Newton's law

$$\mu\frac{dv}{dt} = \frac{\mu}{L} - \mu v\left(2H + \frac{1}{\ell_{\mathrm{f}}}\right)\,. \tag{7}$$

This merely states that curvature accelerates the strings while damping (friction and expansion) slows them down. A relativistic generalization of the velocity evolution equation (7) can be obtained more rigorously by differentiating the definition of $v^2$:

$$\frac{dv}{dt} = \left(1-v^2\right)\left[\frac{k}{L} - v\left(2H + \frac{1}{\ell_{\mathrm{f}}}\right)\right]\,. \tag{8}$$

This involves two assumptions. Firstly, to obtain the damping term we have taken $\langle\dot{\mathbf{x}}^4\rangle = \langle\dot{\mathbf{x}}^2\rangle^2$; the difference between the two is a second-order term. Secondly, in the curvature term, we have introduced $L$ via the definition of the curvature radius vector,

$$\frac{a(\tau)}{L}\hat{\mathbf{u}} = \frac{d^2\mathbf{x}}{ds^2}\,, \tag{9}$$

where $\hat{\mathbf{u}}$ is a unit vector and $s$ is the physical length along the string. The dimensionless parameter $k$ is defined by

$$\langle(1-\dot{\mathbf{x}}^2)(\dot{\mathbf{x}}\cdot\hat{\mathbf{u}})\rangle \equiv kv(1-v^2)\,. \tag{10}$$

Clearly $k$ will be affected by the presence of small-scale structure on strings: on a perfectly smooth string, $\hat{\mathbf{u}}$ and $\dot{\mathbf{x}}$ will be parallel so $k = 1$ (up to a second-order term as above); however this need not be so for a wiggly string. We shall have more to say about $k$ below.

Equations (6) and (8) form the basis of our generalized 'one-scale' model, which we will now proceed to apply in several different contexts. We note that the velocity-independent 'one-scale' model (6) has proved to be successful in describing the large-scale properties of cosmic string networks in numerical simulations. Any deficiencies seem to be associated with the emergence of significant small-scale structure, that is, propagating kinks and wiggles on scales well below $L$. In friction-dominated regimes, therefore, we should anticipate improved quantitative agreement because of the suppression of this substructure.

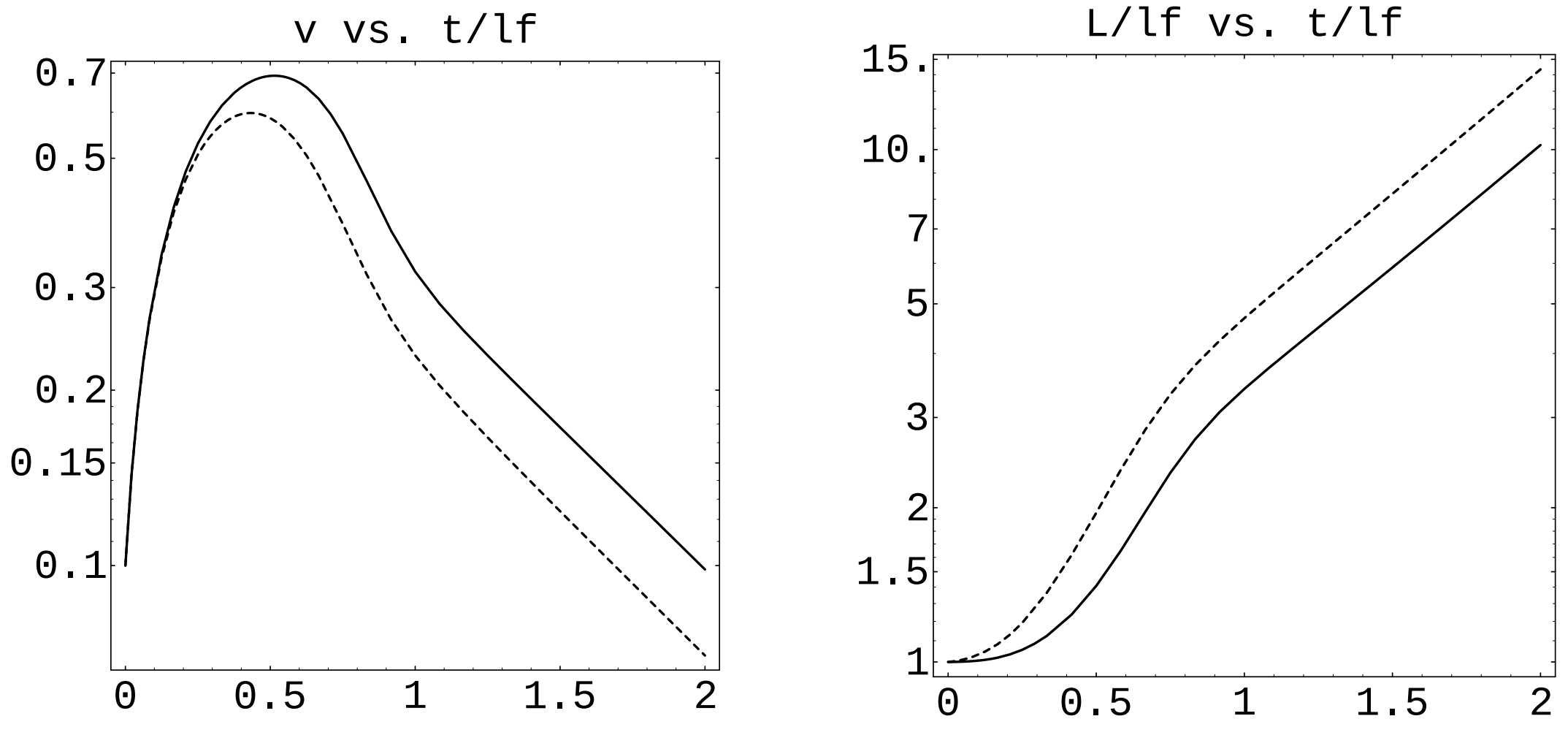

**Figure 1:** Log-log plots of the evolution of $L$ and $v$ in condensed matter systems for $\tilde{c} = 0$ (solid lines) and $\tilde{c} = 1$ (dotted lines), with $k = 1$ and initial conditions $L_i = ct_i = \ell_{\rm f}$, $v_i = 0.1$. The relaxation to scaling is clear.

## 3 Vortices in condensed matter systems

As a simple application, we drop the time-dependence of the damping terms in (6) and (8), that is, we set $H = 0$ and $\ell_{\rm f} = $ const. This should be appropriate for describing vortex–string network evolution in condensed matter systems. The evolution equations become (reintroducing the characteristic speed $c$):

$$2\frac{dL}{dt} = v\left(\frac{L}{\ell_{\rm f}}\frac{v}{c} + \tilde{c}\right), \qquad \frac{dv}{dt} = \left(c^2 - v^2\right)\left(\frac{k}{L} - \frac{v}{c}\frac{1}{\ell_{\rm f}}\right). \tag{11}$$

This yields the following late-time asymptotic behaviour:

$$\frac{L}{\ell_{\rm f}} = \left[k\left(k + \tilde{c}\right)\right]^{1/2}\left(\frac{ct}{\ell_{\rm f}}\right)^{1/2}, \qquad \frac{v}{c} = \left(\frac{k}{k + \tilde{c}}\right)^{1/2}\left(\frac{ct}{\ell_{\rm f}}\right)^{-1/2}. \tag{12}$$

The asymptotic ratio of the loop formation and friction terms is $\tilde{c}/k$; as expected, increasing $\tilde{c}$ (or including loop losses in the first place) leads to a lower network scaling density and a smaller average velocity $v$. Fig. 1 illustrates the relaxation to scaling for a particular set of initial conditions. The duration and nature of the transient regime depends on the initial conditions and the loop 'chopping' efficiency—the larger the value of $\tilde{c}$, the faster the approach to 'scaling'.

The $t^{1/2}$ scaling law for the characteristic lengthscale is a well-known result in the theory of phase ordering with a non-conserved order parameter, supported by simulations and experiment (see [3] for a recent review). In that context, it is obtainde by rather different methods; in particular it should be noted that loop formation is not explicitly included.

## 4 Strings in the early universe

In the early universe the friction lengthscale increases with time (as $t^{3/2}$ in the radiation era), so friction will only be important at 'early times' [6] . Let $T_c$ be the temperature of the string-forming phase transition. In a weakly-coupled Higgs model $G\mu = (T_c/m_{Pl})^2$, the time of string formation is $t_c = (fG\mu)^{-1} m_{Pl}^{-1}$, where $f = 4\pi\left(\pi\mathcal{N}/45\right)^{1/2}$. Defining $t_*$ as the time at which the two damping terms in (2,3) have equal magnitude we find

$$t_* = \frac{\beta^2}{f^3}\frac{m_{Pl}^{-1}}{(G\mu)^2}\,(Rad.)\,, \qquad t_* = \left(\frac{4}{3}\right)^{3/2}\frac{\beta}{f^{3/2}}\frac{m_{Pl}^{-1}}{G\mu}\left(\frac{t_{eq}}{t_{Pl}}\right)^{1/2}\,(Mat.). \tag{13}$$

String dynamics is friction-dominated from $t_c$ until $t_*$, after which motion will become relativistic or 'free'. A simple heuristic argument due to Kibble (see, for example [7]) suggests that in the damped phase the correlation length will scale as $L \propto t^{5/4}$.

As for initial conditions, we assume (following [7]) that $\ell_{\mathrm{f}i} < L_i < t_i \approx t_c$ which, in terms of our variables, is $\tilde{\sigma}^{-1} < \tilde{L}_i < 1$, where $\tilde{\sigma} = \beta/f\,(G\mu)^{1/2}$ and $\tilde{L} = L/t_c$. These two extreme limits could respectively correspond to a rapid second-order phase transition and a slow first-order transition . Note that using the definitions of $\beta$, $\tilde{\sigma}$ and $f$, one finds that the initial ratio of the string and background densities obeys

$$\frac{32\pi}{3}G\mu \le \left(\frac{\rho_\infty}{\rho_b}\right)_i \le \frac{60\zeta(3)}{\pi^4}\omega \quad (\le 0.75\,)\,, \tag{14}$$

(where $0 \le \omega < 1$ is a model-dependent parameter). In fact, analysis of the evolution equations (6,8) shows that for physically reasonable values of $\tilde{c}$ and $k$ these bounds hold for all subsequent times. Consequently, cosmic strings can never dominate the universe.

We now discuss the different evolution histories of electroweak and GUT strings.

**Electroweak strings**

For strings forming around the electroweak phase transition, we have $t_c \approx 4 \times 10^{32} t_{\mathrm{Pl}}$, $t_* \approx 7 \times 10^{25} t_c$ and $\sigma \approx 2 \times 10^{15}$. Note that with $\Omega = 1$ and $h \approx 0.7$ the time of equal matter and radiation densities is $t_{eq} \approx 10^{22} t_c$ (and the present time is $t_0 \approx 10^{28} t_c$), so friction-domination ends well into the matter era. We find that the early time evolution of the string network crucially depends on the initial string density. The complete evolution from $t_c$ through to $t_0$ is illustrated in fig. 2 for several values of $\tilde{L}_i$.

If the initial string density is low, the correlation length is large, so friction dominates over curvature and the strings will move with small velocities. Since both the friction and loop formation terms in (62) are velocity-dependent, we find a period where strings are conformally stretched, with the dynamical variables behaving as

$$\frac{L}{t_c} = \tilde{L}_i\left(\frac{t}{t_c}\right)^{1/2}\,, \qquad v = \frac{k}{\sigma\tilde{L}_i}\frac{t}{t_c}\,. \tag{15}$$

For $\tilde{L}_i$ of order unity (close to the upper limit), this stretching regime can last up to ten orders of magnitude in time. As the velocity increases, the friction and loop formation terms become of the same order of magnitude as the expansion term, and there follows a period where we find Kibble's law

$$\frac{L}{t_c} = \left(\frac{2k(\tilde{c}+k)}{3\sigma}\right)^{1/2} \left(\frac{t}{t_c}\right)^{5/4}, \qquad v = \left(\frac{3k}{2\sigma(\tilde{c}+k)}\right)^{1/2} \left(\frac{t}{t_c}\right)^{1/4}. \tag{16}$$

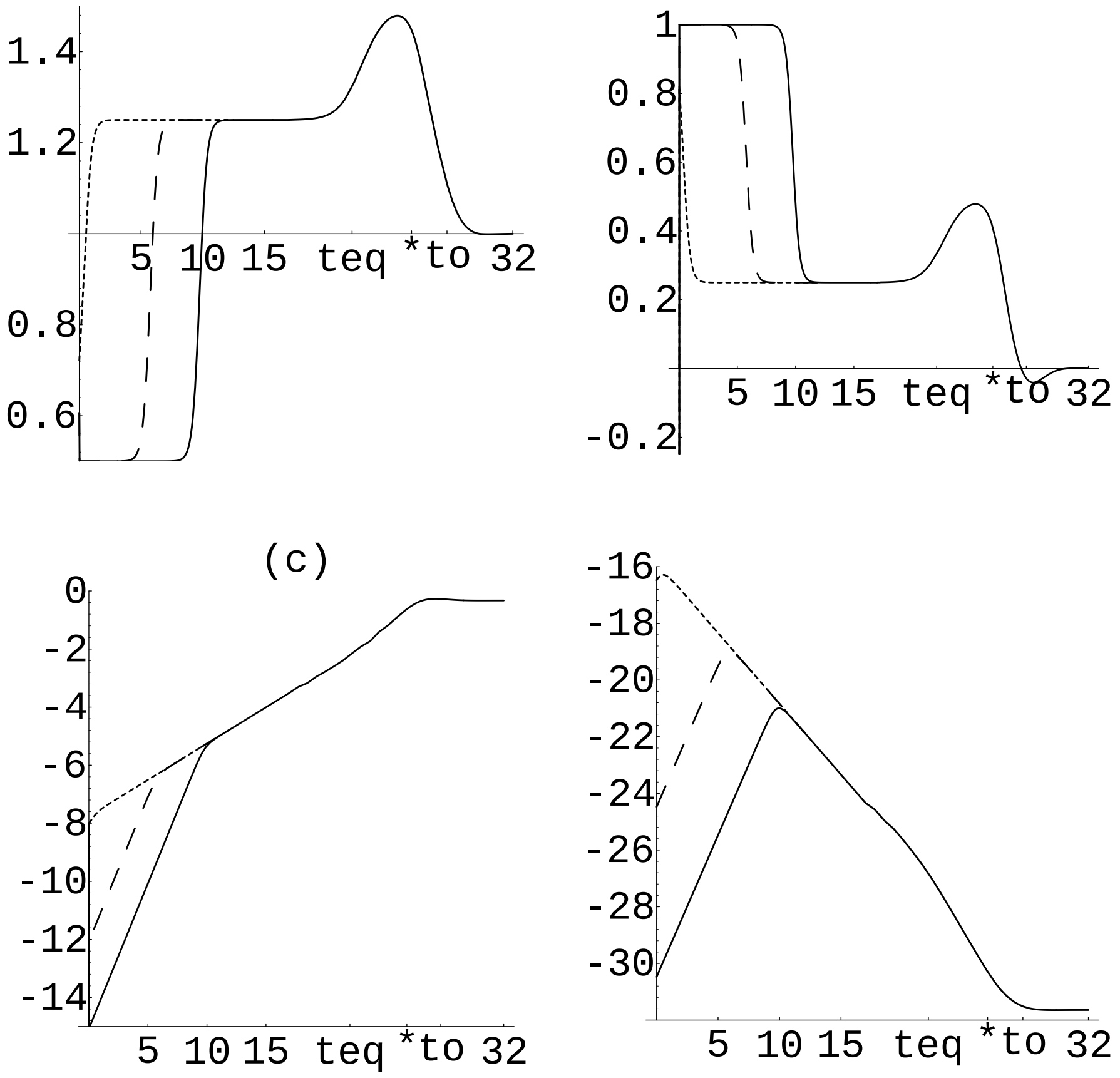

**Figure 2:** The evolution of electroweak string networks. The plots represent the exponent of the power-law dependence of (a) the correlation length $L$ and (b) the average velocity $v$, while (c) shows the actual growth of $v$ and (d) is the ratio of string and background densities (the latter two are log–log plots). Solid, dashed and dotted lines correspond to different initial conditions with $\tilde{L}_i = 10^{-1}$, $\tilde{L}_i = 10^{-4}$ and $\tilde{L}_i = 10^{-8}$, respectively; the horizontal axis is labelled in orders of magnitude in rescaled time $t/t_c$ from the time of formation $t_c$ and also marked are $t_{\rm eq}$, $t_*$ and $t_0$.

However, notice that this only holds in the radiation era; in the corresponding regime in the matter era, we have $L \propto t^{3/2}$ and $v \propto t^{1/2}$. This latter regime is briefly achieved in fig. 2 before friction ceases to dominate the dynamics. The network then evolves towards the final scaling regime:

$$L = \left(\frac{9k(k+\tilde{c})}{8}\right)^{1/2} t, \qquad v = \left(\frac{k}{2(k+\tilde{c})}\right)^{1/2}. \tag{17}$$

Electroweak strings only just reach this 'free' scaling regime by the present day. Note that if we begin with a sufficiently high string density, then there will be no stretching regime (15) and the $L(t) \propto t^{5/4}$ regime will begin immediately. A shortcoming of our model at extremely high densities is the fact that we do not incorporate (otherwise rare) loop reconnections back onto the network.

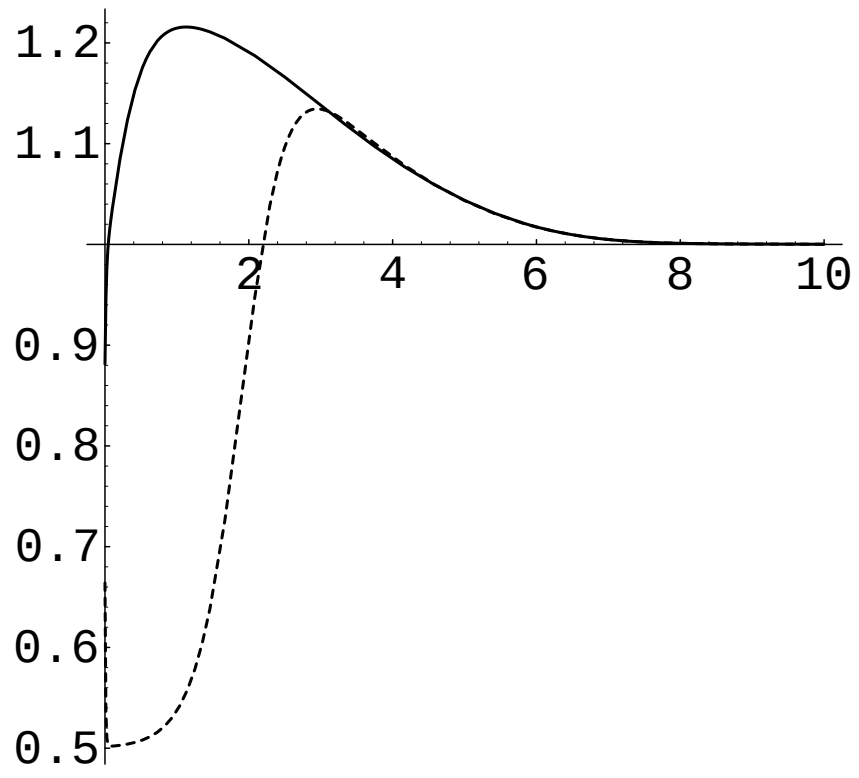

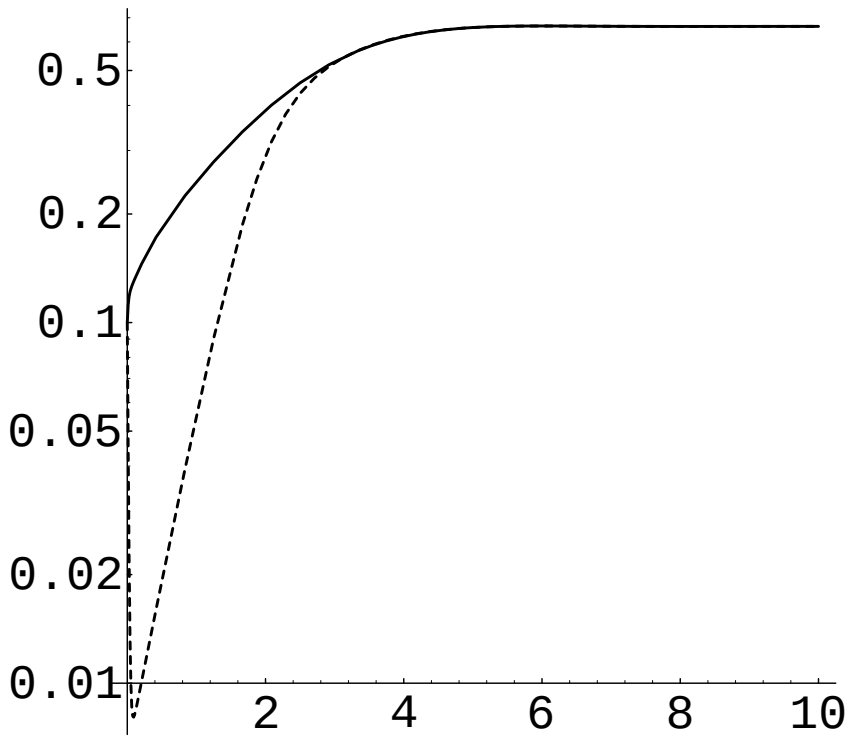

**Figure 3:** The two 'extreme' cases of approach to scaling for minimal-GUT string networks. Solid and dashed lines correspond to high ($\tilde{L}_i = 0.05$) and low ($\tilde{L}_i = 0.9$) initial string density; $\tilde{c}$ and $k$ were chosen to fit radiation-era simulations. Plots represent the exponent of the power-law dependence of $L$ and the average velocity $v$; now $t_\star \approx 2.93$.

**GUT strings**

For a GUT model (with $\mathcal{N} = 106.75$) which produces strings, we have $t_c \approx 3 \times 10^4 t_{Pl}$, $t_* \approx 855 t_c$, $t_{eq} \approx 10^{50} t_c$ and $\sigma \approx 29$. Fig. 3 illustrates the approach to scaling for the two extreme limits for the initial correlation length, $0.035 < \tilde{L}_i < 1$. Since the friction-dominated period between $t_c$ and $t_*$ is much shorter, the nature of the damped evolution becomes much less definite: although we can still identify the 'stretching' and 'friction' regimes, there is no longer a clear $L \propto t^{1/2}$ or $L \propto t^{5/4}$ dependence. In particular, the Kibble regime now depends sensitively on the initial conditions: the higher the initial string density, the higher the maximum exponent in the time-dependence of $L$.

The asymptotic 'free' scaling behaviour in the radiation and matter eras is

$$L = \left(\frac{k(k+\tilde{c})}{e(2-e)}\right)^{1/2} t\,, \qquad v = \left(\frac{(2-e)k}{e(k+\tilde{c})}\right)^{1/2} . \tag{18}$$

For constant $\tilde{c}$ and $k$, this model predicts that the ratios of $\gamma$ and $v_0$ in the matter and radiation eras are independent of $\tilde{c}$ and $k$, being respectively $3/2\sqrt{2}$ and $1/\sqrt{2}$ (whereas simulations, indicate that these ratios are respectively 2 and 0.9 [8]). On the other hand, we can reverse the argument and look for the values of $\tilde{c}$ and $k$ that match the simulations:

$$\tilde{c}_r \approx 0.24\,, \quad k_r \approx 0.18\,, \qquad \tilde{c}_m \approx 0.17\,, \quad k_m \approx 0.49\,. \tag{19}$$

If, as we claim, $k$ is related to the small-scale structure (more small-scale structure corresponding to a smaller $k$) these values are qualitatively in agreement with the simulations. No doubt we require additional degrees of freedom to treat small-scale structure satisfactorily (see, for example, ref. [2]).

The main consequence of our analysis is that minimal-GUT strings approach the linear scaling regime (and, in particular, become relativistic) faster than previously estimated. Another important point concerns the transition between the radiation and matter eras which we find is much slower than previously estimated (extending for about seven orders of magnitude in time, the deviation from the $L \propto t$ dependence never being more than 1%). However we must exercise care in this 'free' regime given the matching problems at $t_{\rm eq}$ for $\tilde{c}$ and $k$.

## 5 Conclusions

In conclusion, we have generalized Kibble's one-scale model by including friction and treating the average long-string velocity as a dynamical variable. We believe this model advances our understanding of the scale-invariant damped evolution of string networks, as well as the relaxation to this 'scaling', in both condensed matter and cosmological contexts. It should form the basis for further progress and, for example, we may be able to gain insight into the effects of string small-scale structure via the parameter $k$ (defined in (10)). In a longer more complete description, we have also applied this model to the study of global strings or vortices (see ref. [9]).

## Acknowledgements

C.M. is funded by JNICT (Portugal) under 'Programa PRAXIS XXI' (grant no. PRAXIS XXI/BD/3321/94). E.P.S. is funded by PPARC and we both acknowledge the support of PPARC and the EPSRC.